\documentclass[10pt]{article}

\usepackage{amsmath}
\usepackage{amssymb}

\usepackage{graphicx}

\usepackage{cite}
\usepackage[normalem]{ulem}

\usepackage{color}

\usepackage[labelfont=bf,labelsep=period,justification=raggedright]{caption}

\makeatletter
\renewcommand{\@biblabel}[1]{\quad#1.}
\makeatother

\date{}

\usepackage{hyperref}

\begin{document}

\begin{flushleft}
{\Large
\textbf{Human Mobility in a Continuum Approach}
}
\\
Filippo Simini$^{1,2,\ast}$, 
Amos Maritan$^{3}$, 
Zolt\'an N\'eda$^{4}$
\\
\bf{1} Center for Complex Network Research and Department of Physics, Biology and Computer Science, Northeastern University, Boston, Massachusetts 02115, USA
\\
\bf{2} Institute of Physics, Budapest University of Technology and Economics, Budafoki \'ut 8, Budapest, H-1111, Hungary
\\
\bf{3} Dipartimento di Fisica e Astronomia ``G. Galilei'', Universit\`a di Padova, CNISM and INFN, via Marzolo 8, 35131 Padova, Italy
\\
\bf{4} Department of Physics, Babe\c{s}-Bolyai University, Kog\u{a}lniceanu street 1, RO-400084, Cluj-Napoca, Romania
\\
$\ast$ E-mail: f.simini@neu.edu
\end{flushleft}

\section*{Abstract}

Human mobility is investigated using a continuum approach that 
allows to calculate the probability to observe a trip to any 
arbitrary region, and the fluxes between any two regions.
The considered description offers a general and unified framework, in which 
previously proposed mobility models like the gravity model, the intervening 
opportunities model, and the recently introduced radiation model are 
naturally resulting as special cases. 
A new form of radiation model is derived and its 
validity is investigated 
using observational data offered by commuting trips obtained from the 
United States census data set, and the mobility fluxes 
extracted from mobile phone data collected in a western European country. 
The new  modeling paradigm offered by this description suggests that the 
complex topological features observed in large mobility and transportation 
networks may be the result of a simple stochastic process taking place 
on an inhomogeneous landscape.

\section*{Introduction}

Human mobility in form of migration or commuting becomes increasingly important nowadays due to 
many obvious reasons \cite{RefWorks:197}: (i)  traveling becomes easier, quicker and 
more affordable; (ii) some borders (like the ones inside EU) are more transparent or even
inexistent for travelers; (iii) the density and growth of the population and their gross national product  presents large territorial inequalities, which naturally induces mobility;
(iv) the main and successful employers  concentrate their location in narrow geographic regions where living costs are high, hence even in developed countries the employees are forced to commute;  (v) large cities grow with higher rates, optimizing their functional efficiency and creating the necessary intellectual and 
economic surplus for sustaining this growth \cite{RefWorks:198}. This higher growth rate of the population can be achieved only by relocating the highly skilled work-force from smaller cities.  
Here we propose a unified continuum approach to explain the resulting mobility patterns.\\
Understanding and modeling the general patterns of human mobility is a long-standing problem in sociology and human geography with obvious impact on business and the economy \cite{RefWorks:199}. 
Research in this area got new perspectives, 
arousing the interest of physicists \cite{RefWorks:94,RefWorks:39} 
due to the availability of several accurate and large scale electronic data, which helps track the mobility fluxes \cite{RefWorks:135,RefWorks:204,RefWorks:179} and check the hypotheses and results of different models. 
Traditionally mobility fluxes were described by models originating from physics. 
The best-known is the {\em gravity model} \cite{RefWorks:126,RefWorks:135} 
that postulates fluxes in analogy with the Newton's law of gravitation, 
where the number of commuters between two locations is proportional 
to their populations (i.e. the `demographic mass') and decays with the square of 
the distance between them.
Beside the well-known gravity model, several other models were used like the generalized potential model
\cite{RefWorks:200,RefWorks:201}, the intervening opportunities model \cite{RefWorks:158}  or the random utility model
\cite{RefWorks:202}. 
Recently, a parameter-free {\em radiation model} has been proposed, leading to mobility patterns 
in good agreement with the empirical observations \cite{RefWorks:192}. The model was developed assuming 
a spatially discretized settlement structure, and consequently it operates with a discretized flux
topology on the edges of a complete graph. Here we consider and test a continuum approach 
to this model operating with fluxes between any two regions, and show that several other mobility models can be derived within the same framework.
This novel approach based on the continuum 
description offers a new modeling and data interpretation paradigm 
for understanding human mobility patterns.\\

\section*{Results}

\subsection*{The modeling framework}

The radiation model \cite{RefWorks:192} has been originally formulated 
to estimate commuting fluxes, i.e. the average number of commuters traveling 
per unit time between any two locations in a country. 
The key idea is that while the home-to-work trip is a daily process, 
it is determined by a one-time choice, i.e. the job selection. 
Therefore commuting fluxes reflect the human behavior in the
choice of the employment. In real life many variables can affect
the employment's choice, from personal aspirations to economic considerations,
but for the sake of simplicity only the most influential variables are
considered in the model: the salary a job pays (or more generally, the working
conditions), and the distance between the job and home. 
The main idea behind
the model is that an individual accepts the closest job with better pay: 
\textit{each individual travels to the nearest location where she/he
can improve her/his current working conditions (benefits)}. 
With this assumption, the probability $P_>(z|a)$ 
that an individual with benefit $z$ refuses the closest $a$ offers is: 
\begin{equation} \label{PC}
	P_>(z|a)  =  p_{\leq}(z)^{a}
\end{equation}
where $a$ is the number of open positions in the area within 
a circle of radius $r(a)$ centered in the origin location, and 
$p_{\leq}(z)=\int_0^z d x\; p(x)$ is the cumulative distribution 
function of the benefits.
Equation~(\ref{PC}) is equivalent with assuming that the rejection of $a$ 
job offers with benefits less or equal to $z$ are independent events.\\
Making different assumptions and approximations on the benefit distribution 
$p(z)$, one can obtain several formulas for the number of trips 
between locations. Below we present four examples: the
{\em original radiation model}, the classic {\em intervening opportunities} (IO) model 
\cite{RefWorks:158}, a {\em uniform selection model}, and a novel 
{\em radiation model with selection}.
\paragraph{The original radiation model.}
If we solve Eq.~(\ref{PC}) assuming that the benefit distribution $p(z)$ 
is a continuous function, we recover the original radiation model's formula \cite{RefWorks:192}. 
Indeed, we calculate the probability $P_>(a)$ of not accepting one of the 
closest $a$ job offers by integrating Eq.~(\ref{PC}) over the benefits:
\begin{eqnarray} 
	P_>(a) &=& \int_0^\infty \mathrm{d} z \frac{d p_{\leq}(z)}{d z} P_>(z|a) \label{Pa}\\
	&=& \int_0^\infty \mathrm{d} z \frac{d p_{\leq}(z)}{d z} p_{\leq}(z)^a \label{PC1final}
	= \frac{1}{a+1}.
\end{eqnarray}

\paragraph{The intervening opportunities model.}
We can also show how the classical IO model \cite{RefWorks:158,RefWorks:217} 
can be included within the same framework as a degenerate case.
Consider the situation in which the benefit distribution is singular, 
i.e. all jobs are exactly equivalent 
$p(z)=\frac{d}{d z}p_{\leq}(z)=\delta(z^*-z)$ 
and 
$p_{\leq}(z)=1 - \Theta(z^*-z)$ (where $\Theta$ is the 
Heaviside function).
In this case we have to specify the individual's behavior when s/he receives a job offer
identical to her/his current one: this corresponds to setting a specific value to the 
step function at the discontinuity point, $\Theta(0)=\kappa$. 
If $\kappa=0$, then the individual will travel to
an infinite distance; while if $\kappa=1$, the individual accepts the
job in the closest location. If $0<\kappa<1$, then the individual
accepts each offer with probability $\kappa$ and refuses it with probability $1-\kappa$.
Applying Eq.~(\ref{Pa}) we obtain
\begin{equation}\label{PC3final}
	P_>(a) = (1-\kappa)^a = \mathrm{e}^{-a L}
\end{equation}
where $L = -\ln(1-\kappa)\cong\kappa$ if $\kappa\cong 0$.
\paragraph{The uniform selection model.}
When $a\cdot L \ll 1$, a good approximation of Eq.~(\ref{PC3final}) is 
$P_>(a) = 1-a L$, which corresponds to randomly select one of the 
available job opportunities, irrespective of the benefits and the distance.
Generalizing this interpretation, we can define a model on a finite 
space containing $N$ average job openings per unit time in which the 
accepted job is selected uniformly at random, and thus $P_>(a) = 1-a/N$.\\
\paragraph{ The radiation model with selection.}
Let us assume that the benefit distribution $p(z)$ is continuous as in the 
original radiation model, whereas the probability  
to accept any offer is reduced by a factor $(1-\lambda)$ with $\lambda\in[0,1]$. 
As a consequence, the probability that an individual with benefit $z$ 
accepts an offer has to be replaced by a reduced value: 
$\tilde{p}_{>}(z) = (1-\lambda)p_{>}(z), \;\forall z>0$.
This process can be interpreted as a commuting population who is willing to accept 
better offers with probability $(1-\lambda)$, or who is aware only 
of a fraction $(1-\lambda)$ of the available job offers. 
This is equivalent to a combination of 
the radiation model and the intervening opportunities model described above 
(here $1-\lambda  = \kappa$).
In this case  
$P_>(z|a) = \tilde{p}_{\leq}(z)^a = [1-\tilde{p}_{>}(z)]^a$, 
and the probability to refuse the closest $a$ offers is 
\begin{align}\label{PC2final}
	P_>(a) &= \int_0^\infty\mathrm{d}z\frac{d p_{\leq}(z)}{dz} [\lambda + (1-\lambda)p_{\leq}(z)]^{a}   =\nonumber\\
	&= \int_0^1 \, \mathrm{d}w [\lambda + (1-\lambda)w]^{a} 
       	= \frac{1-\lambda^{a+1}}{a+1} \frac{1}{(1-\lambda)} 
\end{align}
Note that when $\lambda=0$ we recover the original radiation model (\ref{PC1final}), 
while a $\lambda > 0$ causes a shift of the median of $P_>(a)$ towards 
higher values of $a$. 
In particular, for $\lambda, \lambda' \to 1$ the following approximation holds:  
$P_>(a, \lambda') \approx P_>\left(a \frac{(1 - \lambda')}{(1 - \lambda)}, \lambda \right)$, 
where we made explicit the dependence on $\lambda$. 
The validity of this relationship can be verified by defining $\kappa= 1- \lambda$ 
and expanding around $\kappa \approx 0$: 
\begin{align}\label{median1}
&P_>(a, 1-\kappa') =  \frac{1-(1-\kappa')^{a+1}}{\kappa'(a+1)} \approx  \nonumber\\
&\frac{1-[1-\kappa' (a+1) + \frac{\kappa'^2}{2}  (a+1) a  + \mathcal{O}(\kappa'^3) ] }{\kappa'(a+1)} = 
 1 - \frac{\kappa' a}{2}  + \mathcal{O}(\kappa'^2 a)
\end{align}
and
\begin{align}\label{median2}
&P_>(a \frac{\kappa'}{\kappa}, 1-\kappa) =  \frac{1-(1-\kappa)^{a (\kappa' / \kappa)+1}}{\kappa(a (\kappa' / \kappa) +1)} \approx  \nonumber\\
&\frac{1-[1-\kappa (a (\kappa' / \kappa) +1) + \frac{\kappa^2}{2}  (a (\kappa' / \kappa) +1) a (\kappa' / \kappa)  + \mathcal{O}(\kappa^3) ] }{\kappa(a (\kappa' / \kappa) +1)} =  \nonumber\\
& 1 - \frac{\kappa' a}{2}  + \mathcal{O}(\kappa' \kappa \, a)
\end{align}
The difference is of the order $\mathcal{O}(\kappa'^2 a) - \mathcal{O}(\kappa' \kappa \, a)$, 
thus $|P_>(a, 1-\kappa') - P_>(a \frac{\kappa'}{\kappa}, 1-\kappa)| \to 0$ when $\kappa, \kappa' \to 0$. 
Note that Eq.~(\ref{median2}) follows immediately from Eq.~(\ref{median1}) 
by substituting 
$\kappa' \mapsto \kappa$ and $a \mapsto a \kappa'/\kappa$.
We can derive the dependence of the median on the rescaling of the parameter $\lambda$: 
if with $\lambda = 0.9$ the median is $\tilde a$ defined by $0.5 = P_>(\tilde a | 0.9)$, 
with $\lambda' = 0.99$ the median is ten times higher, i.e. 
$\frac{0.1}{0.01} \tilde a = 10 \, \tilde a$. 
By varying the parameter $\lambda$ it is thus possible to adjust the median of the 
distribution $P_>(a)$, which is equivalent to set a characteristic length of the trips.\\

\begin{figure}[!ht]
\begin{center}
\includegraphics[width=100mm]{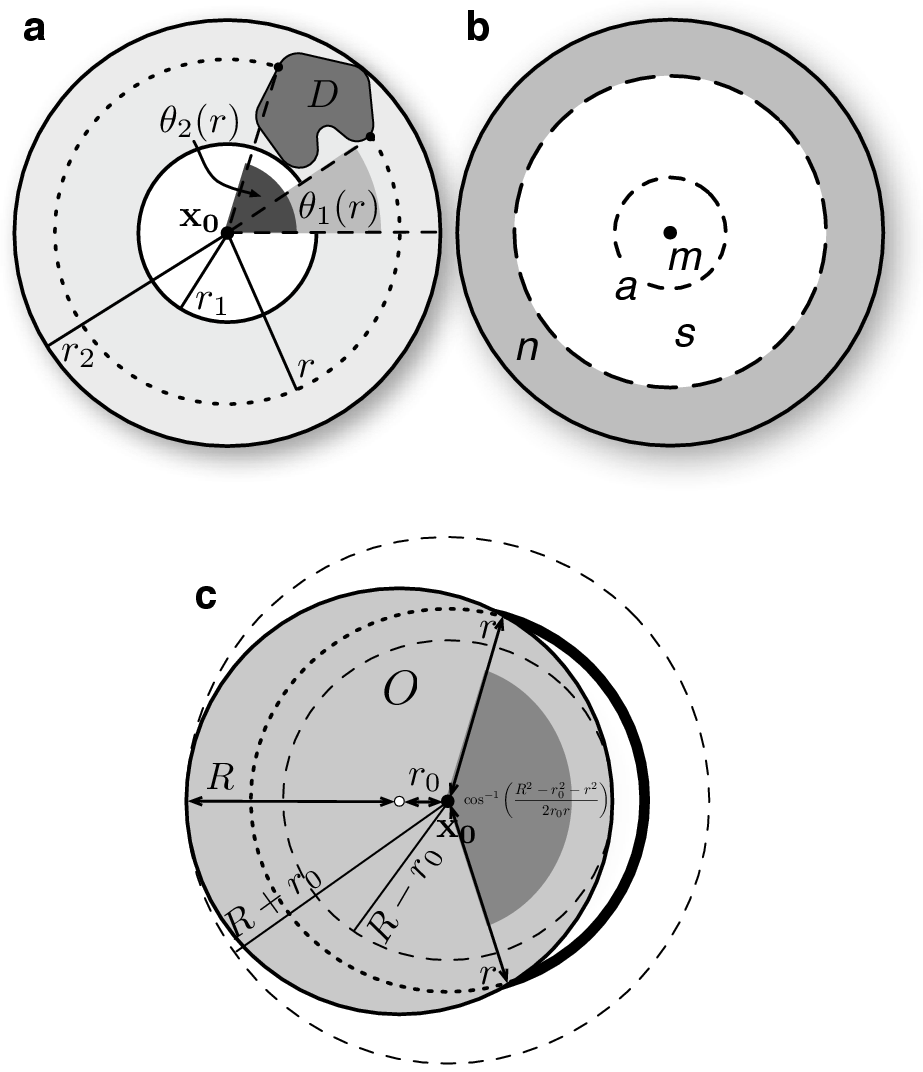}
\end{center}
\caption{
{\bf Definition of the variables used in the calculations.}
a) Notation used in Eq.~\ref{pd}. b) Configuration used to calculate the probability $P(n|a,m)$
c) Configuration used in Eq.~(\ref{TD}) to calculate $\langle T_{O}(O^c)\rangle$.
}
\label{f2}
\end{figure}

These examples show the versatility of the radiation model's formalism, 
which can successfully provide an explanation to several probability 
distributions $P_>(a)$ observed empirically in different contexts
\cite{RefWorks:158,RefWorks:192}. 
The probability density, $P(a)$, to accept one of the offers between $a$ and $a+\mathrm{d} a$ 
for a unit $\mathrm{d} a$ value can
be obtained from $P_>(a)$ by derivation.  To be more specific, let us consider
the original radiation model.  From Eq.~\ref{PC1final} we have $P(a) = -
\frac{d}{d a}P_>(a) = 1/(1+a)^2$.  Let $n(\mathbf {x})$ be the density of
job offers at point $(x,y)$ (in polar coordinate, $(r,\theta)$, we will
use the same notation for the density $n(r,\theta)$). Then one gets the
following expression for the number of job offers within a distance $r$
from $\mathbf{x_0}$,
$a(r)=\int_{|\mathbf{x} - \mathbf{x_0}|\leq r}\mathrm{d} \mathbf{x}\ n(\mathbf
{x})=\int_0^r\mathrm{d}r'\int_0^{2\pi}\mathrm{d}\theta\ r'\
n(r',\theta)$ and
$\mathrm{d}a=r\mathrm{d}r\int_0^{2\pi}\mathrm{d}\theta\ n(r,\theta)$.
Thus the probability to accept an offer within a region at distance
between $r$ and $r+\mathrm{d} r$, $\mathcal{P}(r)\mathrm{d} r$, is given by
\begin{align}\label{px}
\mathcal{P}(r) \mathrm{d} r =P(a)\mathrm{d} a = \frac{\mathrm{d} r}{[1+a(r)]^2} \frac{d a(r)}{d r} = \frac{r\, \mathrm{d} r \int_0^{2\pi} \mathrm{d}\theta\, n(r,\theta)}{[1+a(r)]^2}\  .
\end{align}
This also suggests that 
\begin{equation}\label{pxx}
 P_{\mathbf{x_0}}(\mathbf{x})\mathrm{d}x\mathrm{d}y = \frac{n(\mathbf{x})}{[1+a(|\mathbf{x}|)]^2}\mathrm{d}x\mathrm{d}y
\end{equation}
is the probability to travel from the origin, $\mathbf{x_0}$, to an area $\mathrm{d}x\mathrm{d}y$ centered at the spatial point $\mathbf {x}$.
In general, $P_{\mathbf{x_0}}(\mathbf{x})$ has the following simple expression for any model presented above:
$P_{\mathbf{x_0}}(\mathbf{x}) = - P'_>[a(|\mathbf{x}-\mathbf{x_0}|)] \cdot n(\mathbf{x})$.
From Eq.~(\ref{px}) we can derive the probability $P_{\mathbf{x_0}}(D)$ of a trip 
from the origin to a generic region $D$ (see Fig.~\ref{f2}a) as 
\begin{align}\label{pd}
	P_{\mathbf{x_0}}(D) = \int_D \mathrm{d} \mathbf{x}\, P_{\mathbf{x_0}}(\mathbf{x}) 
        = \int_{r_1}^{r_2}\mathrm{d} r\, \frac{\hat{n}(r)}{[1+a(r)]^2} \frac{\hat{n}_D(r)}{\hat{n}(r)}
\end{align}
where $\hat{n}(r) = r \int_{0}^{2\pi}\mathrm{d}\theta\, n(r,\theta) $
is the radial job offers' density, and 
$\hat{n}_D(r) = r \int_{\theta_1(r)}^{\theta_2(r)}\mathrm{d}\theta\, n(r,\theta)$ 
is the job offers' density in $D$ at distance $r$ from $\mathbf{x_0}$. 
If the radial job offers' density has small variations around its average 
between $r_1$ and $r_2$, i.e.
$\hat{n}_D(r) \approx \frac{1}{r_2-r_1} \int_{r_1}^{r_2}\mathrm{d} r\, \hat{n}_D(r) \equiv \langle\hat{n}_D\rangle_{r_1}^{r_2}$ 
and 
$\hat{n}(r) \approx \langle\hat{n}\rangle_{r_1}^{r_2}$ $\forall r \in [r_1,r_2]$, 
then we can derive a simple approximated formula for $P_{\mathbf{x_0}}(D)$
\begin{align}\label{pd2}
	&P_{\mathbf{x_0}}(D) \overset{(\ref{pd})}{\approx} \int_{r_1}^{r_2}\mathrm{d} r\, \frac{\hat{n}(r)}{a^*(r)^2} 
               \frac{\int_{r_1}^{r_2}\mathrm{d} r\, \hat{n}_D(r)}{\int_{r_1}^{r_2}\mathrm{d} r\, \hat{n}(r)} = \nonumber\\
	      &= \int_{r_1}^{r_2}\mathrm{d} r\, \frac{\frac{d a^*(r)}{d r}}{a^*(r)^2} \frac{a(D)}{a^*(r_2)-a^*(r_1)} = \frac{a(D)}{a^*(r_2) \cdot a^*(r_1)}
\end{align}
where $a^*=1+a$, and $a(D) = \int_D \mathrm{d}\mathbf{x}\, n(\mathbf{x})$ is the number of job offers in $D$.\\
This equation is especially important because data are usually 
collected as fluxes in a discretized space, whose regions are defined according 
to the local administrative subdivision (e.g. counties or municipalities). 
$P_{\mathbf{x_0}}(D)$ has a particularly simple expression if we consider the 
probability $P(n,a)$ to accept one of the $n$ offers between $a$ 
and $a+n$, corresponding to the ring in Fig.~\ref{f2}b. 
This is given by
$P(n,a) = \int_a^{a+n} \mathrm{d} x\, P(x) = P_>(a) - P_>(a+n) = 1/(1+a)-1/(1+a+n)$, 
which in the limit $n\to0$ tends to $n \, P(a)$.
If we only consider trips outside a circular region centered on the 
origin location and containing $m$ job offers, then the 
probability $P(n,a|m)$ to accept one of the $n$ offers between 
$a$ and $a+n$ given that none of the closest $m$ offers has been accepted, is 
$P(n,a) / P_>(m) = \frac{(1+m)n}{(1+a)(1+a+n)}$.
Note that $P(n,a|m)$ is the same probability of one trip 
derived in the original radiation model's discrete formulation \cite{RefWorks:192} with the only difference being that here we have $1+m$ instead of $m$
($a$ is equal to $s+m$).\\
It is important to observe that the equations derived for $P(a)$ 
are correctly normalized when the total number of job offers, $N^{\text{tot}}$, 
is infinite and therefore finite-size corrections 
are required in real-world applications \cite{RefWorks:207}. 
The normalized probability is 
$P(a) / \mathcal{N}$, 
where the normalization constant is 
$\mathcal{N} = \int_0^{N^{\text{tot}}} \mathrm{d}a \, P(a) = P_>(0) - P_>(N^{\text{tot}})$. 
The correction to $P(a)$ is of the order $\mathcal{O}(1 /N^{\text{tot}})$, which in most 
cases is very small given that usually $N^{\text{tot}} \gg 1$. 
This normalization scheme has a straightforward mechanistic interpretation: 
it offers another try at job selection for individuals who during their first 
job search did not find any job offer with better benefit than their current 
one. 
Other kinds of normalization procedures that combine two of the models 
presented above are also possible. 
If , for example, we assume that the individuals who did not find a better job 
in their first try decide to select the offer with the highest benefit, 
even if it does not exceed their current one, (a mechanism corresponding 
to the random selection model) the normalized probability we obtain is 
$P(a) + P_>(N^{\text{tot}}) a / N^{\text{tot}}$. 
Therefore, there are multiple ways to normalize the models, each capturing 
a different selection mechanism. 
This suggests that a systematic investigation of finite size effects could 
also help understand the mechanisms underlying job selection.

\subsection*{Comparison with empirical data}

In Fig.~\ref{f3} we apply the original parameter-free radiation model 
(Eq. \ref{PC1final}) and the one-parameter radiation model with selection (Eq. \ref{PC2final}) 
to commuting data among United States' counties.
We show the agreement between the theoretical
$P(a) = P(a|m) \cdot P_>(m)$ 
distributions and the collapses predicted by the original radiation model, Fig.~\ref{f3}b,
and the radiation model with selection, Fig.~\ref{f3}c.
In Fig.~\ref{f4} we compare the theoretical distributions $P(a)$ 
of the original radiation model, the radiation model with selection, and the IO model, 
to the empirical distributions extracted from a mobile phone database of a 
western European country. 
For a description of the data sets and the analyses performed see 
the section {\em Materials and Methods}.\\
An advantage of the proposed approach is that it is defined for 
a continuous spatial density of job offers, and its results are thus independent 
of any particular space subdivision in discrete locations. 
This feature solves some consistency issues present
in other mobility models defined on a discretized space. 
Consider for example the gravity law \cite{RefWorks:126,RefWorks:134,RefWorks:135}, 
the prevailing framework to predict population movement \cite{RefWorks:120,RefWorks:206,RefWorks:41}, 
cargo shipping volume \cite{RefWorks:124}, inter-city phone calls \cite{RefWorks:127}, 
as well as bilateral trade flows between nations \cite{RefWorks:130}.
The gravity law's probability of one trip from an area with population $m$ 
to an area with population $n$ (assuming that population is proportional 
to the number of job offers) at distance $r$ is obtained by fitting a formula 
like $P(n,r|m)\propto m^\alpha n^\beta f(r)$ to previous mobility data.
As shown in \cite{RefWorks:192}, the values of the best-fit parameters 
$\alpha$ and $\beta$ are strongly dependent on the spatial subdivision considered, 
raising the problem of deciding which subdivision gives the correct results.\\
Also, the continuous formalism developed here helps finding a solution 
to the issue concerning the additivity of the fluxes frequently encountered 
in discrete formulations.
As an example, consider two adjacent areas, $1$ and $2$ with populations $n_1$ 
and $n_2$ respectively, at the same distance $r$ from the origin location. 
The gravity law predicts 
$T(1)=C m^\alpha n_1^\beta f(r)$ and $T(2)=C m^\alpha n_2^\beta f(r)$ 
travelers to $1$ and $2$ respectively.
If we consider a different spatial subdivision, in which locations 1 and 2 are now 
grouped together forming a single location, $1+2$, and we calculate the 
number of travelers we obtain 
$T(1+2) = C m^\alpha (n_1+n_2)^\beta f(r) \neq T(1)+T(2)$
unless $\beta=0$ or $\beta=1$.
If the exponent $\beta$ is different from one, the additivity 
requirement does not hold and the difference in the estimated trips can be
considerably high. For example, if $\beta = 0.5$ and $n_1 = n_2 = 5000$, then
$\Delta T = T(1) + T(2) - T(1+2) \propto 141 - 100 = 41$, i.e. a $41\%$ 
relative difference.
The additivity of the fluxes is a necessary property required to any
mobility model in order to be self-consistent. We can easily verify
that all models derivable from Eq.~(\ref{PC}) have the additivity property.
This is a consequence of the linearity of the integral in Eq.~(\ref{pd}).
In fact, for every two regions $D_1 \cap D_2 = \varnothing$ and $D_1 \cup D_2 = D_{1+2}$
we have 
$\langle T_{\mathbf{x_0}}(1+2) \rangle \propto P_{\mathbf{x_0}}(D_{1+2}) = \left[ \int_{D_1} \mathrm{d}\mathbf{x}\, P_{\mathbf{x_0}}(\mathbf{x})+ \int_{D_2} \mathrm{d} \mathbf{x}\, P_{\mathbf{x_0}}(\mathbf{x}) \right]= \langle T_{\mathbf{x_0}}(1) \rangle + \langle T_{\mathbf{x_0}}(2) \rangle$, 
for a generic $P_{\mathbf{x_0}}(\mathbf{x})$.  
We observe that it is possible to develop a continuum formalism 
for the gravity model
that fulfils the additivity constraint by assuming that the probability to 
travel from location $\mathbf{x_0}$ to location $\mathbf{x}$ is 
$P_{\mathbf{x_0}}^{gm}(\mathbf{x}) = n(\mathbf{x_0})^\alpha n(\mathbf{x})^\beta f(|\mathbf{x}-\mathbf{x_0}|)$. 
The average number of travelers from region $O$ to region $D$ is 
$\langle T_{O}(D)\rangle = \int_O \mathrm{d} \mathbf{x_0} \, n(\mathbf{x_0}) \int_D \mathrm{d} \mathbf{x} \, P_{\mathbf{x_0}}^{gm}(\mathbf{x})$ 
and because of the linearity of the integral on $D$ the fluxes are additive.\\
We can use the continuum approach to investigate the relationship between a region's 
population and the total number of travelers from that region outwards 
(i.e. the commuters whose destination is outside the region). 
It is often assumed that the number of commuters is proportional to the 
region's population. 
This is the case, for example, for the commuting fluxes measured by the 
US census 2000 \cite{RefWorks:192}. 
We can check the validity of this assumption by writing the average number of 
commuters leaving a region $O$ as 
$\langle T_{O}(O^c)\rangle = \int_O \mathrm{d} \mathbf{x_0} \, n(\mathbf{x_0}) P_{\mathbf{x_0}}(O^c) $, 
where $O^c = \mathbb{R}^2 \smallsetminus O$ is the complement of $O$, and 
$P_{\mathbf{x_0}}(O^c) = \int_{O^c} \mathrm{d} \mathbf{x} \,  P_{\mathbf{x_0}}(\mathbf{x})$ 
is the probability for an individual in $\mathbf{x_0}$ to travel outside $O$ (cf. Eq.~\ref{pd}). 
We can easily calculate $\langle T_{O}(O^c)\rangle$ if we make the simplifying assumptions 
that the number of job offers in a region is proportional to the region's population 
(see the section {\em Materials and Methods} for details), that the population density is uniform, 
i.e. $n(\mathbf{x}) = n$, and $O$ is a circle of radius $R$ (see Fig.~\ref{f2}c). 
Then 
\begin{align}\label{TD}
& \langle T_{O}(O^c)\rangle = \int_{O} \mathrm{d} \mathbf{x_0} \, n(\mathbf{x_0}) \int_{O^c} \mathrm{d} \mathbf{x} \, P_{\mathbf{x_0}}(\mathbf{x}) = 
(2 \pi \, n) \int_0^{R} d r_0 \, r_0 \nonumber\\ 
&\left[ \frac{1}{\pi} \int_{R-r_0}^{R+r_0} d r \, \mathcal{P}(r) \cos^{-1}\left(\frac{R^2 - r_0^2 - r^2}{2 r_0 r}\right) + \int_{R+r_0}^{\infty} d r \, \mathcal{P}(r) \right] 
\end{align}
where $\mathcal{P}(r) d r$ is the probability to travel to a distance $r + d r $ 
(cf. Eq.~\ref{px}). 
For the original radiation model $\mathcal{P}(r) = 2 \pi n r / (1 + n \pi r^2)^2$, and Eq.~(\ref{TD}) 
can be calculated exactly and has the following asymptotic limits: 
$\langle T_{O}(O^c)\rangle(R) \sim R^2 n$ if $R \ll n^{-1/2}$, and 
$\langle T_{O}(O^c)\rangle(R) \sim R \sqrt{n} \pi$ if $R \gg n^{-1/2}$. 
The same asymptotic behaviour is obtained for the IO model, with 
$\mathcal{P}(r) = L e^{-L n \pi r^2}$: 
$\langle T_{O}(O^c)\rangle(R) \sim R^2 $ if $R \ll (nL)^{-1/2}$, and 
$\langle T_{O}(O^c)\rangle(R) \sim R $ if $R \gg (nL)^{-1/2}$. 
For both models if the size of the region, $R$, is sufficiently small then 
the number of commuters, $\langle T_{O}(O^c)\rangle(R)$, is proportional to the total 
population of the region. When $R$ becomes larger than a characteristic size 
only the individuals living close to the boundary have a non-zero chance of travelling 
outside $O$.\\
A further generalization of the model could take into account 
the fact that Euclidean distance is not appropriate in situations 
where geographical barriers exist and/or travel facilities are 
heterogeneously distributed. 
In this case one introduces a metric tensor $g_{ij}(\mathbf{x})$ 
and the square distance between neighboring positions at point 
$\mathbf{x}$ is 
$(dr)^2=\sum_{ij}\, g_{ij}(\mathbf{x})\mathrm{d}x_i\mathrm{d}x_j$ 
with $x_1=x$ and $x_2=y$. 
In this case Eq.~(\ref{pxx}) is rewritten as
$P_{\mathbf{x_0}}(\mathbf{x})\mathrm{d}x_1\mathrm{d}x_2 = \frac{n(\mathbf{x})}{[1+a(|\mathbf{x}|)]^2}\sqrt{g(\mathbf{x})}\mathrm{d}x_1\mathrm{d}x_2$, 
where $\sqrt{g(\mathbf{x})}=\det(g_{ij})$ is a local parameter of the model.

\section*{Discussion}

The fundamental Eq.~(\ref{PC}) represents a unified framework to 
model mobility and transportation patterns. 
In particular, we showed how the intervening opportunity model \cite{RefWorks:158}
can be regarded as a degenerate case of the radiation model, corresponding 
to a situation in which the benefit differences are not taken into account
in the employment's choice. We also explained the advantages of a continuous 
approach to model mobility fluxes, we derived the appropriate discretized 
expressions that guarantee
the consistency of our predictions on any discrete 
spatial subdivision, verifying that the fluxes additivity requirement holds.\\
Furthermore, our approach also provides an insight on the theoretical 
foundation of the most common types of gravity models. 
Indeed, when the space is homogeneous and the job's distribution is fractal, $a(r)$ is independent of the point of 
origin, i.e. $a(r) = \rho \, r^{d_F}$ where $d_F$ and $\rho$ are the fractal dimension 
and an average density of job offers, respectively.
Equation (\ref{pd2}) for the probability, $P(D)$, to observe a trip to a generic 
region $D$ within distances $r_1$ and $r_2$ from the origin becomes 
($n=a(D)$ is the number of job offers in D) 
$P(D) \approx [P_>(\rho r_1^{d_F}) - P_>(\rho r_2^{d_F})] \frac{n}{\rho r_2^{d_F} - \rho r_1^{d_F}}$.
In particular, for the original radiation model, Eq. (\ref{PC1final}), 
the average number of trips 
to a region $D$ containing $n$ job offers is
$T(D) \propto \frac{ n}{(\rho^2 r_2 \cdot r_1)^{d_F}}$, 
whereas for the intervening opportunities model, Eq. (\ref{PC3final}), 
$T(D) \propto \frac{ \frac{n} {\rho r_2^{d_F} - \rho r_1^{d_F}}} {\mathrm{e}^{L\rho r_1^{d_F}} - \mathrm{e}^{L\rho r_2^{d_F}}}$.
These two classes of deterrence functions $f(r)$, power law and exponential, 
are actually the two most used form of gravity models \cite{RefWorks:134,RefWorks:41,RefWorks:121}.
Moreover, our approach provides an interpretation to the 
gravity model's fitting parameters. 
First, the exponents $\alpha$ and $\beta$ are both one when the benefits are 
spatially uncorrelated, i.e. the benefit distributions at the local (regional) 
and global (country) scales are the same. 
If $\alpha$ or $\beta$ differ from one it means that there are regions where 
job offerings with higher or lower benefits tend to concentrate.
Second, the exponent of the power law is predicted to be two times the 
fractal dimension of the job offers, $d_F$,
whereas the exponential deterrence 
function should be substituted with a stretched exponential with shape parameter
$d_F$ and a characteristic length of the order of $(\rho L)^{-1/d_F}$.
Thus, when the spatial displacement of the potential trip's destinations 
is a fractal, the radiation model's formalism offers a theoretical derivation 
of the gravity models from first principles.\\
In conclusion, we have developed a general framework for 
unifying the  theoretical foundation of a broad class of human mobility models. 
The used continuum approach allows for a consistent description of mobility 
fluxes between any delimited regions. 
The successful comparison with real  
mobility fluxes extracted from two different data sources 
confirms that our approach not only provides a theoretically sound modeling 
framework, but also a good quantitative agreement with experimental data.
This suggests that the decision process we assumed for the job selection  
also captures the basic decision mechanism related to the choice of  
the destinations for other activities (shopping, leisure, ...). 
On the other hand, our study suggests that the weighted network representing 
the mobility fluxes among geographic regions can be the 
result of a stochastic process consisting of many independent 
events.
This approach is somehow complementary to the theory of optimal 
transportation networks \cite{RefWorks:143,RefWorks:144,RefWorks:193,RefWorks:205,RefWorks:194,RefWorks:195} 
that describes the patterns 
observed in different natural and artificial systems solely 
as the adaptation to a global optimization principle
(e.g. leaf venations, river networks, power grids, 
road and airport networks).
The modeling framework we propose provides also a plausible 
example of spontaneous bottom-up design of transportation networks. 
Indeed, we show how complex patterns can arise even in those 
systems lacking a global control on the network topology, or a 
long-term evolutionary selection mechanism of the optimal structure.\\

\section*{Materials and Methods}

\subsection*{Analysis on the inter-county commuting trips extracted from United States' Census data}

\begin{figure}[ht!]
\begin{center}
\includegraphics[width=160mm]{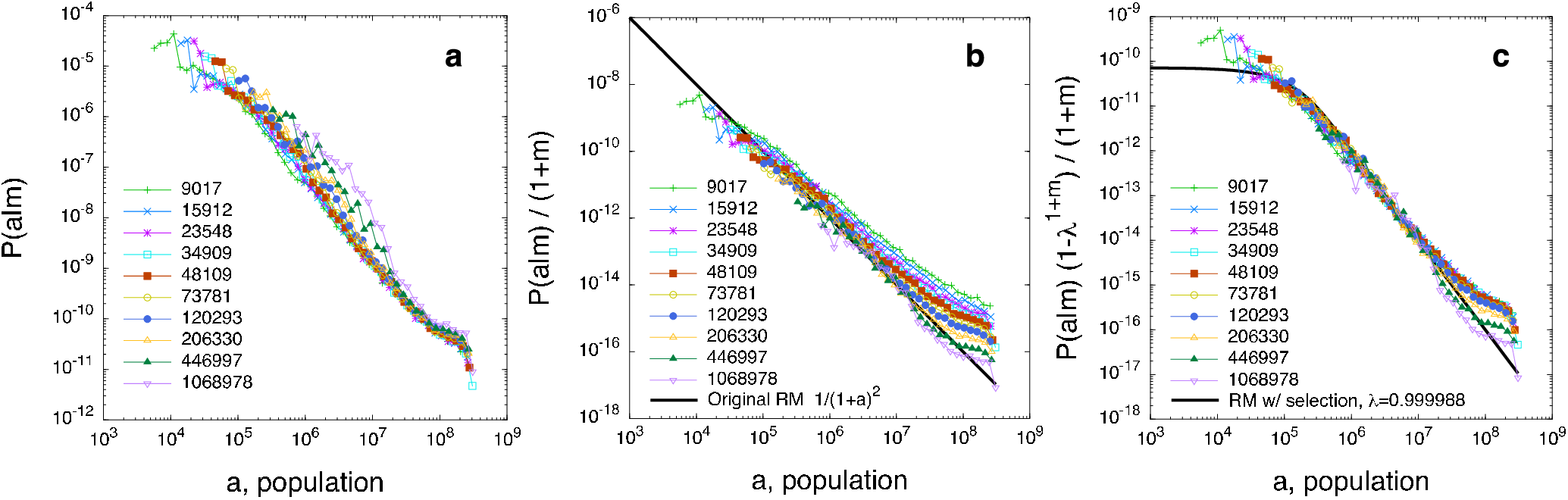}
\end{center}
\caption{
{\bf Testing the radiation model's theoretical predictions on 
commuting trips extracted from the US census dataset.}
\emph{a}) We divide the commuting flows in deciles according to 
the population of the origin county, $m$, and for each set we calculate the distributions 
$P(a|m)$. The values in the key indicate the mean origin population, $m$, of each decile. 
We use the population as a proxy to estimate the number of employment opportunities in 
every county, $a$, assuming in first approximation a linear relationship 
between population and job openings.
\emph{b},\emph{c}) The collapse of the distributions $P(a)=P(a|m)\cdot P_>(m)$ 
on the theoretical curves Eqs.~(\ref{PC1final}) and (\ref{PC2final})
predicted by the original radiation model and the radiation model with selection respectively.
(See the section {\em Materials and Methods} for details).
}
\label{f3}
\end{figure}

The data on US commuting trips can be freely downloaded from\\ 
{\em http://www.census.gov/population/www/cen2000/commuting/index.html}.\\ 
The files were compiled from Census 2000 responses to the long-form (sample) questions on
where individuals worked, and provide all the work destinations for people who live in each county.
The data contain information on 34,116,820 commuters in 3,141 counties. \\
Demographic data containing the population and the geographic coordinates 
of the centroids of each county can be freely downloaded from 
{\em https://www.census.gov/geo/www/gazetteer/places2k.html}.\\ 
Our goal is to use the US commuting data to calculate the empirical distribution 
$P(a) = -\frac{d}{d a}P_>(a)$ and compare it to the theoretical predictions of the 
original radiation model, Eq.~(\ref{PC1final}), and the radiation model with selection, 
Eq.~(\ref{PC2final}).\\ 
We assume that the number of employment opportunities in every county, $a_{\text{jobs}}$, 
is proportional to the county's population, $a_{\text{pop}}$, i.e.  
$a_{\text{jobs}} =  a_{\text{pop}} \cdot c$, where $c<1$ is the ratio 
between the average number of job offers considered by an individual 
(i.e. the ones known and of potential interest) over the population. 
Under this assumption, if we calculate the probability $P(a)$ using the 
population instead of the job openings the resulting distribution is 
simply rescaled as $P(a_{\text{jobs}} /c) / c$.\\ 
From the census data we obtain the fraction of individuals who live in county $i$
with population $m$ and work in county $j$ that lies beyond a circle containing 
a population $a$ as 
$P(a) = T_{ij} / m = (T_{ij} / T_i) (T_i / m) = P(a|m)\cdot P_>(m)$, 
where $T_{ij}$ is the number of commuters from $i$ to $j$, and $T_i$ is the 
total number of commuters from $i$ to all other counties. 
It follows that upon rescaling with $P_>(m)$, all the $P(a|m)$ should collapse 
on the theoretical distribution $P(a)$. 
This is what we want to test in Fig.~\ref{f3}.\\ 
First, we divide the commuting fluxes in deciles according to the population 
of the origin county, $m$. 
Then, for each set we calculate the distributions $P(a|m)$ (Fig.~\ref{f3}a), 
and the rescaled distributions $P(a)=P(a|m)\cdot P_>(m)$ with $m$ equal to 
the mean origin population of the counties in each set, and using the $P_>(m)$ 
of Eq.~(\ref{PC1final}) in Fig.~\ref{f3}b, and of Eq.~(\ref{PC2final}) 
in Fig.~\ref{f3}c.
The value of the parameter $\lambda=0.999988$ has been obtained 
by maximizing the likelihood that the observed fluxes are an outcome of the model.
The discrepancy observed at very high $a$ ($\gg 10^7$) can be the result of boundary 
(finite-size) effects that become relevant at large populations, corresponding to long distances.
Also, the fluctuations at very small $a$ values are due to the resolution 
limit encountered when $a \approx m$. 
The parameter $\lambda$ is close to 1 because in the comparison with data 
we consider populations instead of job offers and we assume that the 
two quantities are proportional, and consequently the fitting parameter 
we find is $\lambda_{\text{pop}} = \lambda_{\text{jobs}}^c$, which is 
always close to 1 irrespective of $\lambda_{\text{jobs}}$ given that $c \ll 1$.\\

\subsection*{Analysis on trips extracted from a mobile phone dataset.} 

\begin{figure}
\begin{center}
\includegraphics[width=85mm]{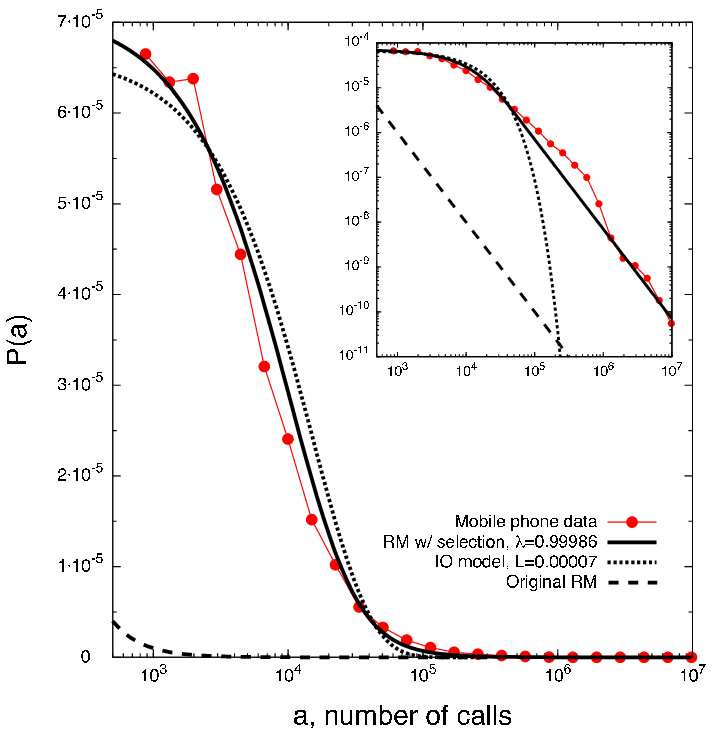}
\end{center}
\caption{
{\bf Testing the mobility models on trips extracted from a mobile phone dataset.}
We analyze all call records collected during one day, and we define a trip when 
we observe two consecutive calls by the same user from two different towers. 
We define the variable $a(D)$, representing the number of possible points of interest 
in a circular area $D$ centered at a given cell tower, as the total number of calls 
placed from the towers in $D$, assuming that a location's attractiveness is 
proportional to its call activity. 
We then calculate the empirical distribution $P(a)\mathrm{d}a$, i.e. the fraction of 
trips to the towers between $a$ and $a+\text{d}a$ (red circles), and we compare it 
to the various models' theoretical predictions $P(a)=-\frac{d}{d a} P_>(a)$, 
with $P_>(a)$ defined in Eqs.~(\ref{PC2final}), (\ref{PC3final}), and (\ref{PC1final}),
and whose parameters, $\lambda=0.99986$ and $L=0.00007$, are obtained with least-squares 
fits (black lines).
In the inset we show the plot in a log-log scale.
(See the section {\em Materials and Methods} for details).
}
\label{f4}
\end{figure}

We use a set of anonymized billing records from a European mobile
phone service provider \cite{RefWorks:131,RefWorks:39,RefWorks:46}. 
The dataset contains the spatio-temporal information of the calls placed by 
$\sim 10 $M anonymous users, specifying date, time and the cellular antenna (tower) that handled each call.
Coupled with a dataset containing the locations (latitude and
longitude) of cellular towers, we have the approximate location of the caller
when placing the call.
We analyze all call records collected during one day, and we define a trip when 
we observe two consecutive calls by the same user from two different towers. 
The type of mobility information obtained from the mobile phone data is 
radically different from that provided by the census data.
In fact, the scope and method of the mobile phone data collection 
is complementary to the self-reported information of the census survey, 
and it offers the possibility to consider all trips, not only commuting 
(home-to-work) trips.
Additionally, the mobility information that we extract from the mobile 
phone data is more detailed in both time and space.
Indeed, we can observe trips of any duration, ranging from few minutes to several hours.
In a similar manner, we can analyse trips on the much finer spatial resolution of  
cellular towers, whose average distance is $\sim 1$km, compared to the 
average size of counties, $\sim 10$km.
We are therefore including in the current analysis many more trips, obtaining 
a more complete picture of individual mobility.\\ 
In Figure \ref{f4} we use the trips obtained from the mobile phone data to 
provide a direct test of the models' fundamental prediction, i.e. the 
specific functional form of the trips distribution $P(a)$. 
In the case of mobile phone data the trips' destinations are determined 
by the particular purpose of the users when they start the trip. 
Therefore, the variable $a$ should now represent not only the number of 
job opportunities in a region, but rather the number of all possible 
venues that could be the destination of a trip, e.g. 
shopping centers, restaurants, schools, bars, etc.  
We therefore define the variable $a(D)$, representing the number of possible points of interest 
in a circular region $D$ centered at a given cell tower, as the total number of calls 
placed from the towers in $D$, assuming that a location's attractiveness is 
proportional to its call activity. 
We then calculate the empirical density distribution $P(a)\mathrm{d}a$, i.e. the fraction of 
trips to the towers between $a$ and $a+\text{d}a$, and we compare it 
to the various models' theoretical predictions $P(a)=-\frac{d}{d a} P_>(a)$, 
with $P_>(a)$ defined in Eqs.~(\ref{PC2final}), (\ref{PC3final}), and (\ref{PC1final}),
and whose parameters, $\lambda=0.99986$ and $L=0.00007$, are obtained with least-squares fits.
Moreover, we verified (plots not shown) that the result presented 
in Fig.~\ref{f4} is stable with respect to other possible ways of defining 
a trip using the mobile phone data, e.g. between the two farthest locations 
visited by each user in 24 hours, or between the two most visited locations. 

\section*{Acknowledgments}

The present work was supported by research grant PN-II-ID-PCE-2011-3-0348. 
AM research is supported by the Cariparo foundation and Prin.
The research leading to these results has received funding from the 
European Union Seventh Framework Programme (FP7/2007-2013) under 
grant agreement No. 270833.
We thank J.~P.~Bagrow, A.-L.~Barab\'asi, F.~Giannotti, J.~S.~Juul, and D.~Pedreschi for many useful discussions.\\

\section*{Figure Legends}

\begin{thebibliography}{10}
\providecommand{\url}[1]{\texttt{#1}}
\providecommand{\urlprefix}{URL }
\expandafter\ifx\csname urlstyle\endcsname\relax
  \providecommand{\doi}[1]{doi:\discretionary{}{}{}#1}\else
  \providecommand{\doi}{doi:\discretionary{}{}{}\begingroup
  \urlstyle{rm}\Url}\fi
\providecommand{\bibAnnoteFile}[1]{%
  \IfFileExists{#1}{\begin{quotation}\noindent\textsc{Key:} #1\\
  \textsc{Annotation:}\ \input{#1}\end{quotation}}{}}
\providecommand{\bibAnnote}[2]{%
  \begin{quotation}\noindent\textsc{Key:} #1\\
  \textsc{Annotation:}\ #2\end{quotation}}
\providecommand{\eprint}[2][]{\url{#2}}

\bibitem{RefWorks:197}
Cohen JE, Roig M, Reuman DC, GoGwilt C (2008) International migration beyond
  gravity: A statistical model for use in population projections.
\newblock Proceedings of the National Academy of Sciences 105: 15269.
\input{RefWorks:197}

\bibitem{RefWorks:198}
Bettencourt L, West G (2010) A unified theory of urban living.
\newblock Nature 467: 912-913.
\input{RefWorks:198}

\bibitem{RefWorks:199}
Ritchey PN (1976) Explanations of migration.
\newblock Annual review of sociology 2: 363-404.
\input{RefWorks:199}

\bibitem{RefWorks:94}
Brockmann D, Hufnagel L, Geisel T (2006) The scaling laws of human travel.
\newblock Nature 439: 462-465.
\input{RefWorks:94}

\bibitem{RefWorks:39}
Gonz\'alez MC, Hidalgo CA, Barab\'asi AL (2008) Understanding individual human
  mobility patterns.
\newblock Nature 453: 779-782.
\input{RefWorks:39}

\bibitem{RefWorks:135}
Barth\'elemy M (2010) Spatial networks.
\newblock Physics Reports 499: 1-101.
\input{RefWorks:135}

\bibitem{RefWorks:204}
Bazzani A, Giorgini B, Rambaldi S, Gallotti R, Giovannini L (2010) Statistical
  laws in urban mobility from microscopic gps data in the area of florence.
\newblock Journal of Statistical Mechanics: Theory and Experiment 2010: P05001.
\input{RefWorks:204}

\bibitem{RefWorks:179}
Eubank S, Guclu H, Kumar VSA, Marathe MV, Srinivasan A, et~al. (2004) Modelling
  disease outbreaks in realistic urban social networks.
\newblock Nature 429: 180-184.
\input{RefWorks:179}

\bibitem{RefWorks:126}
Zipf GK (1946) The p 1 p 2/d hypothesis: On the intercity movement of persons.
\newblock American Sociological Review 11: 677-686.
\input{RefWorks:126}

\bibitem{RefWorks:200}
Anderson TR (1956) Potential models and the spatial distribution of population.
\newblock Papers in Regional Science 2: 175-182.
\input{RefWorks:200}

\bibitem{RefWorks:201}
Lukermann F, Porter PW (1960) Gravity and potential models in economic
  geography.
\newblock Annals of the Association of American Geographers 50: 493-504.
\input{RefWorks:201}

\bibitem{RefWorks:158}
Stouffer SA (1940) Intervening opportunities: a theory relating mobility and
  distance.
\newblock American Sociological Review 5: 845-867.
\input{RefWorks:158}

\bibitem{RefWorks:202}
Block H, Marschak J (1960) Random orderings and stochastic theories of
  responses.
\newblock Contributions to probability and statistics 2: 97-132.
\input{RefWorks:202}

\bibitem{RefWorks:192}
Simini F, Gonz\'alez MC, Maritan A, Barab\'asi AL (2012) A universal model for
  mobility and migration patterns.
\newblock Nature 484: 96.
\input{RefWorks:192}

\bibitem{RefWorks:217}
Schmitt RR, Greene DL (1978) An alternative derivation of the intervening
  opportunities model.
\newblock Geographical Analysis 10: 73-77.
\input{RefWorks:217}

\bibitem{RefWorks:207}
Masucci A, Serras J, Johansson A, Batty M (2012) Gravity vs radiation model: on
  the importance of scale and heterogeneity in commuting flows.
\newblock Arxiv preprint arXiv:12065735 .
\input{RefWorks:207}

\bibitem{RefWorks:134}
Erlander S, Stewart NF (1990) The gravity model in transportation analysis:
  theory and extensions.
\newblock Vsp.
\input{RefWorks:134}

\bibitem{RefWorks:120}
Jung WS, Wang F, Stanley HE (2008) Gravity model in the korean highway.
\newblock EPL (Europhysics Letters) 81: 48005.
\input{RefWorks:120}

\bibitem{RefWorks:206}
Eggo RM, Cauchemez S, Ferguson NM (2011) Spatial dynamics of the 1918 influenza
  pandemic in england, wales and the united states.
\newblock Journal of The Royal Society Interface 8: 233-243.
\input{RefWorks:206}

\bibitem{RefWorks:41}
Viboud C, Bjornstad ON, Smith DL, Simonsen L, Miller MA, et~al. (2006)
  Synchrony, waves, and spatial hierarchies in the spread of influenza.
\newblock Science 312: 447-451.
\input{RefWorks:41}

\bibitem{RefWorks:124}
Kaluza P, K\"olzsch A, Gastner MT, Blasius B (2010) The complex network of
  global cargo ship movements.
\newblock Journal of The Royal Society Interface .
\input{RefWorks:124}

\bibitem{RefWorks:127}
Krings G, Calabrese F, Ratti C, Blondel VD (2009) Urban gravity: a model for
  inter-city telecommunication flows.
\newblock Journal of Statistical Mechanics: Theory and Experiment 2009: L07003.
\input{RefWorks:127}

\bibitem{RefWorks:130}
P\"oyh\"onen P (1963) A tentative model for the volume of trade between
  countries.
\newblock Weltwirtschaftliches Archiv : 93-100.
\input{RefWorks:130}

\bibitem{RefWorks:121}
Balcan D, Colizza V, Gon\c{c}alves B, Hu H, Ramasco JJ, et~al. (2009)
  Multiscale mobility networks and the spatial spreading of infectious
  diseases.
\newblock Proceedings of the National Academy of Sciences 106: 21484.
\input{RefWorks:121}

\bibitem{RefWorks:143}
Li G, Reis SDS, Moreira AA, Havlin S, Stanley HE, et~al. (2010) Towards design
  principles for optimal transport networks.
\newblock Phys Rev Lett 104: 018701.
\input{RefWorks:143}

\bibitem{RefWorks:144}
Hu Y, Wang Y, Li D, Havlin S, Di Z (2011) Possible origin of efficient
  navigation in small worlds.
\newblock Physical Review Letters 106: 108701.
\input{RefWorks:144}

\bibitem{RefWorks:193}
Bohn S, Magnasco MO (2007) Structure, scaling, and phase transition in the
  optimal transport network.
\newblock Physical Review Letters 98: 88702.
\input{RefWorks:193}

\bibitem{RefWorks:205}
Caldarelli G (2007) Scale-Free Networks: Complex webs in nature and technology.
\newblock Oxford University Press, USA.
\input{RefWorks:205}

\bibitem{RefWorks:194}
Durand M (2007) Structure of optimal transport networks subject to a global
  constraint.
\newblock Physical Review Letters 98: 88701.
\input{RefWorks:194}

\bibitem{RefWorks:195}
Corson F (2010) Fluctuations and redundancy in optimal transport networks.
\newblock Physical Review Letters 104: 48703.
\input{RefWorks:195}

\bibitem{RefWorks:131}
Song C, Qu Z, Blumm N, Barab\'asi AL (2010) Limits of predictability in human
  mobility.
\newblock Science 327: 1018.
\input{RefWorks:131}

\bibitem{RefWorks:46}
Onnela JP, Saram\"aki J, Hyv\"onen J, Szab\'o G, Lazer D, et~al. (2007)
  Structure and tie strengths in mobile communication networks.
\newblock Proceedings of the National Academy of Sciences 104: 7332.
\input{RefWorks:46}

\end{thebibliography}
\end{document}